\documentclass[runningheads]{llncs}
\usepackage{graphicx}
\usepackage{algorithm}% http://ctan.org/pkg/algorithms
\usepackage{algpseudocode}% http://ctan.org/pkg/algorithmicx

\begin{document}
\title{Robustification of Segmentation Models Against Adversarial Perturbations In Medical Imaging}
\author{Hanwool Park\inst{1} \and
Amirhossein Bayat\inst{1,2} \and Mohammad Sabokrou \inst{3} \and Jan S. Kirschke \inst{2} \and Bjoern H. Menze \inst{1}} 
% index{Park, Hanwool}
% index{Bayat, Amirhossein}
% index{Sabokrou, Mohammad}
% index{Kirschke, Jan S.}
% index{Menze, Bjoern H.}

% \author{*}
%
% \authorrunning{H. Park and A. Bayat}

%\institute{Department of Informatics, Technical University of Munich, Germany \and
%Department of Neuroradiology, Klinikum rechts der Isar, Germany}

 \institute{Department of Informatics, Technical University of Munich, Germany \and Department of Neuroradiology, Klinikum rechts der Isar, Germany \and School of Computer Science, IPM Institute For Research In Fundamental Sciences, Iran\\
 \email{zerg468@gmail.com}}
%  \\ \email{amir.bayat@tum.de}}

\maketitle              % typeset the header of the contribution
\begin{abstract}
This paper presents a novel yet efficient defense framework for segmentation models against adversarial attacks in medical imaging. In contrary to the defense methods against adversarial attacks for classification models which widely are investigated, such defense methods for segmentation models has been less explored. Our proposed method can be used for any deep learning models without revising the target deep learning models, as well as can be independent of adversarial attacks. Our framework consists of a frequency domain converter, a detector, and a reformer. The frequency domain converter helps the detector detects adversarial examples by using a frame domain of an image. The reformer helps target models to predict more precisely. We have experiments to empirically show that our proposed method has a better performance compared to the existing defense method.

\keywords{Deep Learning  \and Adversarial attacks \and Image segmentation \and Medical imaging}
\end{abstract}

\section{Introduction}
\label{section:Intro}

Recently, Deep learning plays an important role on medical industry since the deep learning system makes it possible to help medical imaging segmentation to identify the pixels of different lesions or organs in medical imaging such as MRI images. However, many researchers discovered that the deep learning neural networks are vulnerable to adversarial examples[3, 11].
To defend against the adversarial attacks, numerous defense algorithms and methods have been introduced and can be grouped under different approaches: (1) training the target deep learning neural networks with the adversarial examples to be more robust; namely adversarial training[3, 11] 
(2) changing training procedures of the target deep learning models for the attackers hard to attack by reducing the magnitude of gradient, e.g.,defensive distillation\cite{Distillation} and (3) building a defense framework, which is independent of the adversarial attacks, to defend the target deep learning models against any adversarial attacks[5, 17]. 
However, All of these approaches have limitations. (1) needs the specific adversarial examples to train the models, so it is only resistant against the limited adversarial attacks. For (2), this approach does not show a good defensive performance against some of the adversarial attacks. (3) shows good generalization and robustness against any adversarial attacks, but some of the defense frameworks only prove their capacities against the small size of images and it is still a challenging task to appropriately train and tune for some of the defense frameworks.

The related works for our proposed defense method are defense approach (3). This approach does not need to modify the target models, as well as it does not need to know any knowledge of the process for crafting adversarial examples. MagNet\cite{MagNet} and Defense-GAN\cite{Defense} are known for this approach. 
MagNet is created to defend the neural network classifiers against the adversarial examples in 2017. MagNet consists of a detector and a reformer. The detector discerns between normal images and adversarial images by measuring how far the input images differ from the manifold of the normal images. The reformer pushes the adversarial images close to the manifold of the legitimate images to help target models to correctly predict. MagNet is the attack-independent defense framework but it works only with the small size of images and classifier models. 
Dense-GAN was introduced by Samangoeui et al. in 2018. This defense framework leverages Generative Adversarial Networks(GANs)\cite{GAN} to defend the deep learning models against any adversarial attacks. Defense-GAN utilizes GANs for the main architecture as well as Gradient Descent(GD) minimization to find latent codes for GANs instead of using the detector network and the reformer network. The authors believed if GAN is trained properly and has enough capacity to represent the data, its reconstructions and the original images should not defer much. However, it is very challenging to train GANs and if GANs are neither trained nor tuned appropriately, the performance of Defense-GAN decreases.

In this paper, we introduce a new defense mechanism to defend the deep learning segmentation models against the adversarial examples in medical imaging. Our approach does not require the target networks to modify the model architectures or training procedures. Moreover, it does not need to know the information about the adversarial examples. Additionally, we use frequency domain of an image instead of spatial domain because we can analyze the geometric characteristic of an image and show more clearly differences between a normal image and an adversarial image in the frequency domain of image because of the difference between low and high frequency domain information. Our contributions are as followed. Firstly, we show that our method can be robust and generalized well against the adversarial examples for semantic segmentation tasks in medical imaging. Secondly, we utilize a frequency domain of an image to differentiate the legitimate examples and the adversarial examples. Thirdly, our defense mechanism can be used for any deep learning neural network models.

\section{Methodology}
\label{section:Mth}

\begin{figure}[h]
  \centering
  \includegraphics[width=0.7\textwidth]{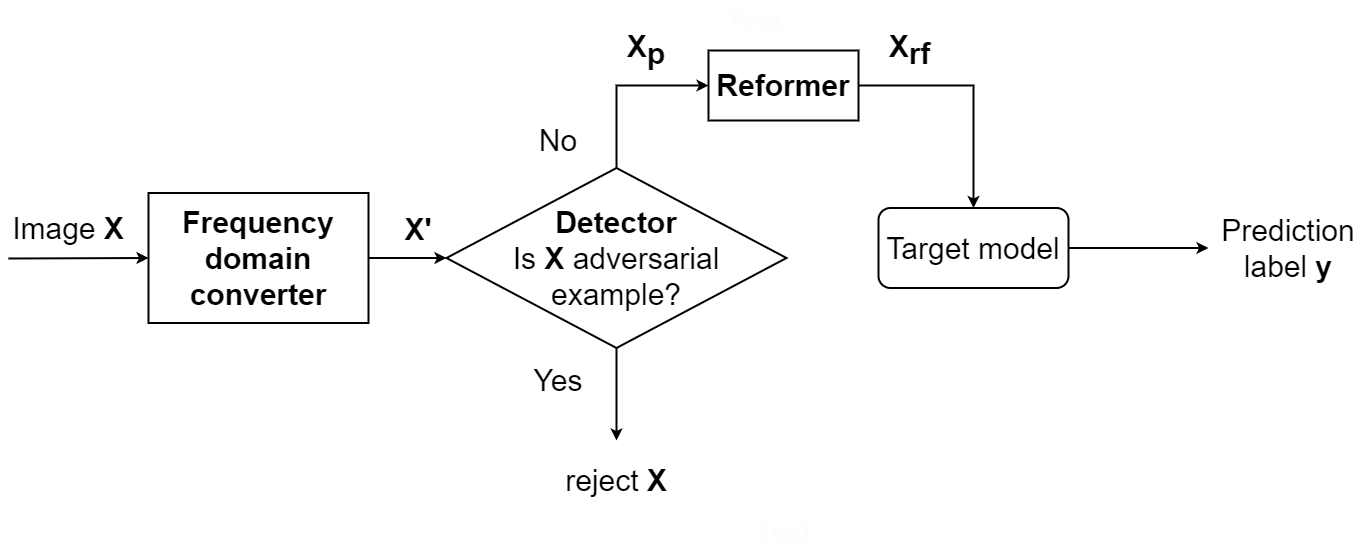}
  
  \caption{The workflow of our defense strategy in the test phase}%{Numerically solved solution}
  \label{fig:ourApproach_overview}
\end{figure}

The purpose of our defense strategy is to design the defense mechanism to be robust and generalized well across any adversarial attacks. Our defense strategy is similar to MagNet. The main differences between MagNet and our approach are : (1) we exploit the deep semantic segmentation models(e.g. UNet\cite{UNet}, DenseNet\cite{DenseNet}) instead of auto-encoders, (2) we utilize a frequency domain of an image to detect the adversarial examples rather than a spatial domain of an image. The main components in our defense strategy are the frequency domain converter, the detector network, and the reformer network. Figure \ref{fig:ourApproach_overview} demonstrates the workflow of our approach in the test phase.

At the test time, the input image $X$ is given to the frequency domain converter at first. The role of the frequency domain converter is to convert the spatial domain of an image into the frequency domain of an image. The detector gets frequency domain of $X'$ which is the output of the frequency domain converter and then differentiates whether $X'$ is the normal image or the adversarial image by measuring the reconstruction error between $X'$ and the manifold of the normal examples, which are also in the frequency domain. If $X'$ is far from the majority boundary(i.e. the reconstruction error is large), the detector network rejects $X'$. The output of the detector is the information of indices of $X'$ which is not rejected by the detector network. We extract the passed input image $X_p$ by using the output of the detector. Then, the reformer pushes $X_p$ close to the manifold of an original image. The important point is that the reformer does not use the frequency domain of an image for its function, but uses the spatial domain of an image. At the final step, the output of the reformer network $X_{rf}$ is given as the input of the target deep segmentation models. Our defense framework only needs these three components. Therefore, we do not need the target networks to change training procedures as well as the model architecture. Moreover, we do not require the information about the adversarial examples.

\medskip

\textbf{Frequency Domain Converter.} The frequency domain converter is a function to return the frequency domain of an image from the spatial domain of an image. The purpose of this function is to help the detector to distinguish between the normal examples and the adversarial examples more precisely. The frequency domain is a space that represents the sine and cosine components of images by using the Fourier Transform\cite{FourierTransform}. In the frequency domain, each point in domain shows a particular frequency in the spatial domain of an image. To be more specific, high-frequency components correspond to edges or boundaries in an image, and low-frequency components represent smooth regions in an image. Hence, we can analyze the geometric characteristics of an image by the frequency domain of the image. The different images have their own frequency domains. Each frequency domain has their dominating directions, which are represented the regular patterns in the background of the images. But, all of the frequency domains have one thing in common. Their magnitudes get smaller for higher frequencies. Therefore, more image information is at low frequencies than at high frequencies. Since the frequency domain represents the rate at which the pixel values are changing in an image, we assume that the frequency domain shows more clearly differences between the normal examples and the adversarial examples than the spatial domain does.

\begin{figure}[!h]
  \centering
  \includegraphics[width=0.5\textwidth]{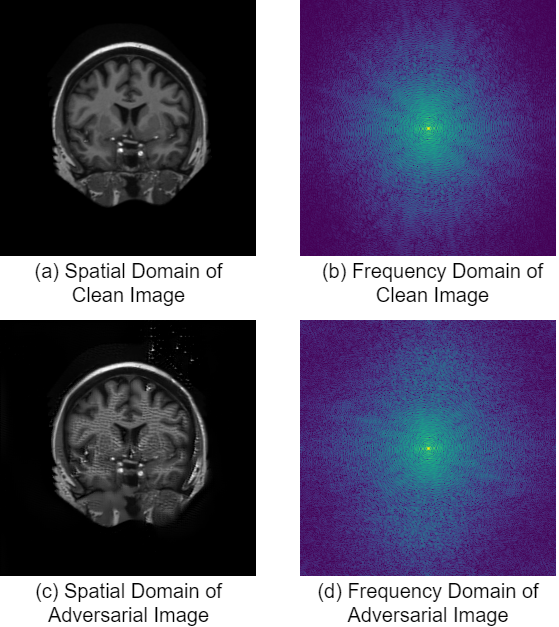}
  \caption{The spatial domain and the frequency domain of one clean image and its corresponding adversarial image}%{Numerically solved solution}
  \label{fig:Fre_Spa}
\end{figure}

The difference between the spatial domain and the frequency domain of the normal image and its corresponding adversarial example is as shown in Figure \ref{fig:Fre_Spa}. Figure \ref{fig:Fre_Spa} (b) represents the frequency domain of the clean image, whereas Figure \ref{fig:Fre_Spa} (d) represents frequency domain of the adversarial image crafted by DAG\cite{DAG}. There are a couple of differences between (b) and (d). Firstly, the dominating directions which are shown in (b) are blurred in (d). Also, (d) has more magnitude for high frequencies than (b). We assume that adding small perturbations to the clean image increases edges and breaks the regular patterns of the clean image. Therefore, we believe that this characteristic of the frequency domain makes the detector to differentiate the adversarial examples more precise compared to the detector using the spatial domain.

\begin{figure}[!h]
  \centering
  \includegraphics[width=0.95\textwidth]{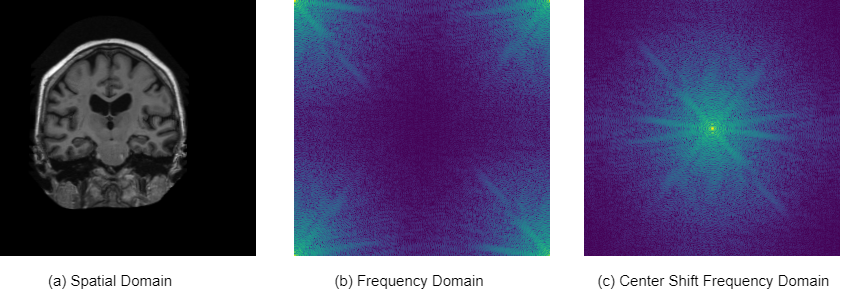}
  \caption{A whole-brain MRI image in the spatial domain, the frequency domain and the frequency domain using the shift function}%{Numerically solved solution}
  \label{fig:shift_fft}
\end{figure}

To convert image from spatial domain to frequency domain, we use Discrete Fourier Transform(DFT)\cite{DFT} because images are discrete signal.The DFT is defined as:

\[ F(u,v) = \frac{1}{WH}\sum_{x = 0}^{W-1}\sum_{y=0}^{H-1}f(x,y)e^{-j2\pi (ux/W + vy/H)} \]

where F(u,v) is the coefficient of the periodic function component with frequency u in the x-axis and v in the y-axis and f(x,y) is the coefficient of the image with the size W x H. The ranges of each component are $x = 0, 1, ... , W - 1$, $y = 0, 1, ... , H - 1$, $u = 0, 1, ... , W - 1$, and $v = 0, 1, ... , H - 1$. 

However, the zero-frequency component(0,0) of the frequency domain of an image by DFT is not at the center of the image as shown in Figure \ref{fig:shift_fft}. It is hard to understand the shape of the frequency domain with (b) since the magnitude becomes bigger close to edges. If we want better to understand the frequency domain of an image, it is better to shift the zero-frequency component to the center of the image. Because F(u,v) is the periodic function that is symmetric with the origin, we can use the shift function to shift the zero-frequency component. We evaluate the performance between the frequency domain detector without using the shift function and with using the shift function in experiments.

\medskip

\textbf{Detector.} The detector is a function that classifies whether the input image is the adversarial image or the legitimate image. We utilize the distance metrics to access the distance between the input examples and the manifold of the normal examples. we train the detector networks with mean squared error(MSE):

\[MSE(X) = \frac{1}{|X|} \sum_{x \in X} \left \| x - D(x)  \right \|_{2} \]

We use a reconstruction error for our distance metrics. $D$ is referred as the detector. The reconstruction error RE(x) on test data x is defined:

\[ RE(x) = \left \| x - D(x) \right \|_{p} \]

If the input sample is drawn from the same data generation process as the training samples which are used for training the detector, the reconstruction error is small. Otherwise, the reconstruction error is big. Hence, we can utilize the reconstruction error for our distance metrics to tell how far a test image is from the majority of the normal images. Also, it is very essential to choose suitable norms when we calculate the reconstruction errors because the norm has a big impacts on the sharpness of detection results. Smaller $p$ in $p$-norm averages its depth to each pixel. On the other hand, a larger $p$ is more responsive to the maximum difference among all pixels. Empirically, we found out that 2-norm is sufficient. Hence, we use 2-norm to calculate the reconstruction error for the detector network. 

Moreover, we must set a threshold $t_{re}$ to decide the decision boundary to detect the adversarial examples since the reconstruction error is a continuous value. Hence, the threshold is a hyperparameter which is an instance of the detector network. The one thing to note is that we should care to select the correct threshold. To detect little perturbed adversarial examples, the threshold should be as low as possible. If the threshold is too low, the detector classifies the normal examples as the adversarial examples. To choose $t_{re}$ , we exploit the validation dataset including the normal examples. We calculate all of the thresholds on the validation dataset and then we discover the highest $t_{re}$ below the threshold $t_{fp}$, which is the false positive rate of the detector on the validation dataset. $t_{fp}$ should be chosen by the requirement of the system. We give one example of $t_{fp}$ and $t_{re}$ to explain the relationship between them. If the $t_{fp}$ is set to 0.1 by the system, the detector rejects no more than 10 \% examples on the validation dataset. So, the threshold $t_{re}$ is the highest value in the threshold set which are already computed by using the validation dataset. 

In MagNet, they implemented the autoencoder architecture for the detector network. However, we assume that the autoencoder appropriately works only with the small size images because the autoencoder architecture is simple network. Hence, we utilize more complicated architectures than autoencoder architectures to have better the detector network. For this, we implement the deep image segmentation models such as UNet and DenseNet. 

\medskip

\textbf{Reformer.} The reformer is a function to try to reconstruct the input image to push close to the majority of the normal examples. The output of the reformer is then given as the input to the target deep learning neural network model. The one thing is to note that the reformer is used only at the inference time. We train the reformer network to minimize the reconstruction error on the training dataset including the legitimate images. Since the reformer is trained with the normal examples, the output of the reformer should be very similar when the input image is reconstructed in the same process as the training normal examples. However, when the the adversarial example comes to the reformer, the reformer moves the adversarial example closer to the majority of the normal examples. Therefore, the reformer can help the target deep neural network model to improve the accuracy of the prediction of the adversarial examples while keeping the accuracy of the prediction of the normal examples. 

%\IEEEpubidadjcol

\section{Experiments}

The following sections describe experiments and results to evaluate the performance of our defense strategy outlined in section \ref{section:Mth}. We introduce our setup for the experiment. Then, we evaluate the performance of the detector networks, as well as we compare our defense approach and MagNet under the adversarial attacks.

\medskip

\textbf{Setup.} Since we are motivated by the paper where Paschali. M. et al., evaluated the performance of SegNet\cite{SegNet}, UNet and DenseNet in unseen clean and adversarial examples\cite{Generalization}, we also select the same segmentation models(SegNet, UNet and DenseNet) for our target models to evaluate the performance of our defense method against the adversarial examples. SegNet does not have skip connection, UNet has long-range skip connections and DenseNet has long-range and short-range skip connections. In the paper \cite{Generalization}, DenseNet shows the best robustness results against adversarial attacks and noisy attacks because of long-range and short-range skip connections followed by UNet and SegNet. 

The semantic segmentation models are trained with the special loss function\cite{Error} combined with weighted multi-class logistic loss and Dice loss \cite{loss}. The loss function is

\[ L = -\sum_{x}w(x)g_{l}(x)log(p_{l}(x))) - \frac{2\sum_{x}p_{l}(x)g_{l}(x)}{\sum_{x}p^2_{l}(x) + \sum_{x}g^2_{l}(x)} \]

where the estimated probability of pixel $x$ belonging to class $l$ represents $p_{l}(x)$ and the its actual class is  $g_{l}(x)$. The first term is the weighted multi-class logistic loss and the second term is the Dice loss. The details are described in the paper written by Roy. AG. et al.\cite{Error}. Model optimization is performed with ADAM optimizer with an initial learning rate of 0.001. 

For our experiments, we use special MRI data from the publicly-available whole-brain segmentation benchmark which is a subset of Open Access Series of Imaging Studies(OASIS)\cite{OASIS}. OASIS dataset has 35 different volumes. The OASIS dataset consists of three-dimensional images, but we split the dataset into two-dimensional images. The adversarial examples are crafted for all of the trained segmentation models by using DAG. To measure the performance of the semantic segmentation models, we use Dice score \cite{Dicescore}. 

To compare our defense framework and MagNet, we implement two segmentation models(UNet, DenseNet) for our detector and reformer. For the detector, we use the training dataset that is trained for the semantic segmentation models. But, we transform the training dataset from the spatial domain to the frequency domain. For the reformer, we train the training examples in the spatial domain. The optimizer is ADAM which is set to the initial learning rate of 0.001 and we use $L^2$ regularization.

\medskip

% https://towardsdatascience.com/understanding-auc-roc-curve-68b2303cc9c5
% https://developers.google.com/machine-learning/crash-course/classification/roc-and-auc

\textbf{Evaluation for Detection.} We evaluate the performance of the different detector architectures. The purpose of this experiment is not only to compare the auto-encoder architectures and the segmentation architectures that we implement for our detector network, but also to compare the detector based on the frequency domain and the detector based on the spatial domain. For this, we utilize the Receiver Operating Characteristics(ROC) - Area Under The Curve(AUC) curve \cite{ROC}.

For this experiment, we design the new OASIS dataset including the normal examples and all types of the adversarial examples crafted by DAG that sets for the three trained segmentation models. The ratio of each examples is the normal examples (50 \%), DAG adversarial examples (50 \%).  

\begin{table}[!h]
\centering
\caption{The ROC-AUC score of each detector network}
\resizebox{0.7\textwidth}{!}{%
\begin{tabular}{c|c|c|c|c|}
\cline{2-5}
 & \multicolumn{4}{c|}{ROC\_AUC score \%} \\ \hline
\multicolumn{1}{|c|}{} & UNet & SegNet & DenseNet & Average \\ \hline
\multicolumn{1}{|c|}{Autoencoder I} & 54.27 & 57.87 & 57.18 & 56.44 \\ \hline
\multicolumn{1}{|c|}{Autoencoder II} & 53.17 & 56.09 & 56.10 & 55.12 \\ \hline
\multicolumn{1}{|c|}{UNet\_spatial} & 84.58 & 97.36 & 92.49 & 91.48 \\ \hline
\multicolumn{1}{|c|}{DenseNet\_spatial} & 80.29 & 95.21 & 87.56 & 87.69 \\ \hline
\multicolumn{1}{|c|}{UNet\_frequency} & 52.74 & 53.81 & 52.90 & 53.15 \\ \hline
\multicolumn{1}{|c|}{DenseNet\_frequency} & 52.39 & 55.35 & 55.49 & 54.41 \\ \hline
\multicolumn{1}{|c|}{UNet\_shiftFrequency} & \textbf{96.12} & \textbf{98.82} & \textbf{99.34} & \textbf{98.09} \\ \hline
\multicolumn{1}{|c|}{DenseNet\_shiftFrequency} & 90.62 & 98.09 & 97.42 & 95.38 \\ \hline
\end{tabular}%
}
\label{Tab:ROC}
\end{table}

Table \ref{Tab:ROC} shows the ROC-AUC score of the different detectors.  Autoencoder I and autoencoder II for MagNet show that they are not capable of discerning the adversarial examples. Our methods with the segmentation models in the spatial domain demonstrate the good performance of classification. Now, we observe the detectors based on the frequency domain. The UNet and DenseNet frequency domain detectors without using the shift function show bad capacity to distinguish between the adversarial examples and the legitimate examples. However, the frequency domain detectors using the shift function show very good measurement of separability

As a result, this experiment provides empirical pieces of evidence to our assumptions. Firstly, deep segmentation models achieve better performance than auto-encoders because auto-encoders cannot differentiate the big size of adversarial images and normal images. Lastly, the detector can differentiate the normal examples and the adversarial examples better when the detector uses the frequency domain instead of the spatial domain.

\medskip

\textbf{Overall Performance against Adversarial Attacks.} We evaluate different defense architectures including MagNet and our defense method against the DAG adversarial attacks. Firstly, we introduce the different defense architectures for the detectors and the reformers. Then, we evaluate the different defense architectures for the trained segmentation models using the DAG adversarial attacks.

% \usepackage{graphicx}
\begin{table}[!h]
\caption{The Combination of Detector and Reformer networks}
\centering
\resizebox{0.8\textwidth}{!}{%
\begin{tabular}{|c|c|c|c|c|c|}
\hline
 & Detector & Reformer &  & Detector & Reformer \\ \hline
1 & Autoencoder I  & Autoencoder I  & 9 & UNet\_frequency & UNet \\ \hline
2 & Autoencoder I & Autoencoder II & 10 & UNet\_frequency & DenseNet \\ \hline
3 & Autoencoder II & Autoencoder I & 11 & DenseNet\_frequency & UNet \\ \hline
4 & Autoencoder II & Autoencoder II & 12 & DenseNet\_frequency & DenseNet \\ \hline
5 & Autoencoder I & UNet & 13 & UNet\_shiftFrequency & UNet \\ \hline
6 & Autoencoder I & DenseNet & 14 & UNet\_shiftFrequency & DenseNet \\ \hline
7 & Autoencoder II & UNet & 15 & DenseNet\_shiftFrequency & UNet \\ \hline
8 & Autoencoder II & DenseNet & 16 & DenseNet\_shiftFrequency & DenseNet \\ \hline
\end{tabular}%
}
\label{Tab:Comb}
\end{table}

Table \ref{Tab:Comb} depicts all the combinations of the detector and the reformer networks for experiments. Diversified defense architectures boost the defense performance. The total number of defense architectures is 16. The combinations of the defense architectures are between the number 1 and the number 4 for MagNet. To compare the performance of the reformers between the auto-encoders and the segmentation models, we design the defense architectures which include the auto-encoders for the detector and the segmentation models for reformer in the number 5 to the number 8. The number 9 to the number 16, we design the frequency domain detectors using UNet and DenseNet. The difference between "frequency" and "shiftFrequency" is that the "shiftFrequency" detectors use the shift function. In this experiment, we do not use the detectors using the segmentation models for the spatial domain because our main idea of the defense framework uses the detector for the frequency domain rather than the spatial domain, as well as we already empirically proved that the frequency domain detector has better power to detect adversarial examples than the spatial domain detector.

Since we have the 16 different defense architectures, we can evaluate the performances of the reformer networks between the auto-encoders and the segmentation models as well as the performances of the detectors between the spatial domain and the frequency domain. Lastly, we can ultimately measure the performance of the defense strategies between our defense method and MagNet. For convenience, we refer to the combination of the defense architectures between the number 1 and the number 16 as between (1) and (16). 

\begin{figure}[!h]
  \centering
  \includegraphics[width=1\textwidth]{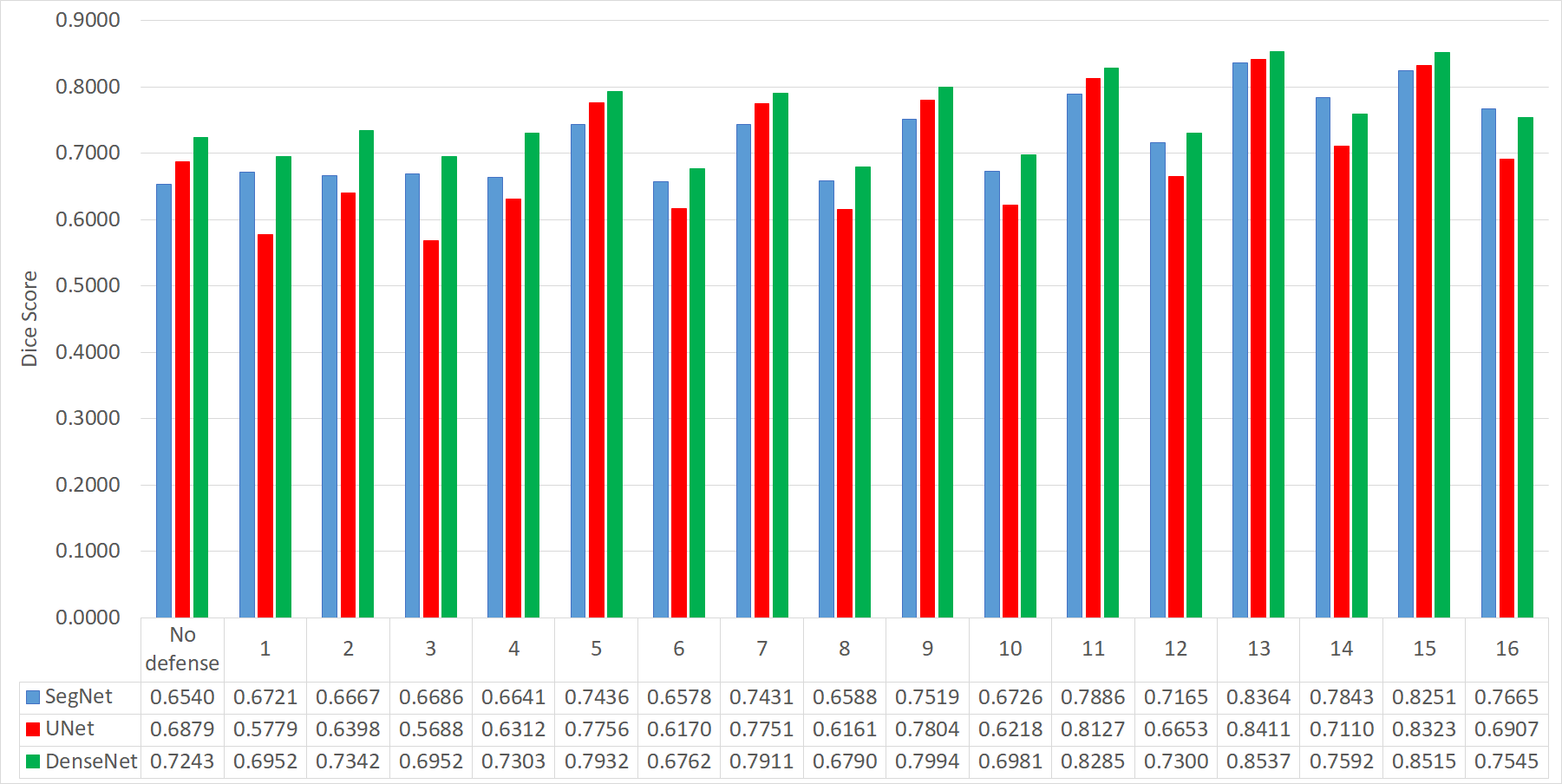}
  \caption{Bar charts for the average Dice scores of SegNet,UNet and DenseNet using the different defense strategies against all of the attacks}%{Numerically solved solution}
  \label{fig:avg_defense}
\end{figure}

Figure \ref{fig:avg_defense} demonstrates the overall performances of the diversified defense architectures for the semantic segmentation models(SegNet, UNet, DenseNet) with the average score against all of the DAG adversarial attacks. We choose the false positive rate of the detector on the validation set $t_{fp}$ to 0.05 to set the threshold of reconstruction error $t_{re}$. 

%The bar charts have two different data labels. The blue label shows the Dice score of different defense strategies at $t_{fp} = 0.01$.

In Figure \ref{fig:avg_defense}, we can compare the performance of each semantic segmentation model without defense. DenseNet shows the best dice score against DAG adversarial attacks followed by UNet and SegNet. Therefore, we can say that the skip connections help to be robust against adversarial attacks. 

Firstly, we look at the bar charts of SegNet with the different defense strategies. MagNet((1) $\sim$ (4)) slightly improves the performance rather than no defense SegNet. Compared with the reformer networks between the auto-encoders((1) $\sim$ (4)) and the segmentation models, the DenseNet reformers((6),(7)) show the same or less performances than the auto-encoders, but the performance of the UNet reformers((5),(7)) improves around 10 \% better than the auto-encoders. Compared with the frequency domain detectors((9) $\sim$ (16)), the UNet shift frequency domain detectors((13),(14)) achieve the better performances than the UNet no shift frequency domain detectors((9),(10)). The DenseNet shift frequency domain detectors((15),(16)) also shows the better achievements than the DenseNet no shift frequency domain detectors((11),(12)). Furthermore, we can compare the detectors between the auto-encoders((5) $\sim$ (8)) and the segmentation models using the shift function in the frequency domain((13) $\sim$ (16)). We observe that the shift frequency domain detectors have better achievements than the auto-encoders detectors. In conclusion, the shift frequency domain detector using UNet and the UNet reformer network (13) shows the best Dice score accuracy compared with the other defense strategies followed by (15). Furthermore, our defense mechanism(13) achieves better than MagNet in SegNet.

Secondly, we observe the bar charts of UNet. MagNet((1) $\sim$ (4)) shows the worse average Dice score than no defense UNet. The combinations with detectors using the auto-encoders and the DenseNet reformers((6), (8)) also display that they are not able to defend against the adversarial examples. The defense strategies with the shift frequency domain detectors and the DenseNet reformers((10), (12)) show the bad performances too. We only observe that the defense strategies using the UNet reformers((5), (7), (9), (11), (13), (15)) are capable of defending against the adversarial attacks. The reformers using UNet architecture ((5), (7)) show better achievements 17 \% than the reformers using the auto-encoders((1) $\sim$ (4)). The shift frequency domains((13) $\sim$ (16)) achieve more than 7 \% than than the auto-encoders detectors((5) $\sim$ (8)). The best performance of the defense strategy is the shift frequency domain detector using UNet and the UNet reformer(13). This strategy shows around 23 \% better performance than MagNet(2).

Next, we observe the overall performances of DenseNet using the different defense combinations. DenseNet also shows that MagNet ((1) $\sim$ (4)) is not very useful to defend against the adversarial attacks. The defense strategies using the UNet reformer((5), (7), (9), (11), (13), (15)) achieve better Dice scores than the defense strategies using the DenseNet reformer((6), (8), (10), (12), (14), (16)). The shift frequency domain detectors((13) $\sim$ (16)) show better achievements than the auto-encoder detectors((5) $\sim$ (8)). The reformer using segmentation models((5) $\sim$ (8)) are more powerful to defend facing the adversarial attacks compared with the reformer using the auto-encoders((1) $\sim$ (4)). The defense strategy(13) shows the best performance among the other defense strategies and the other semantic segmentation models.

In conclusion, these experiments result provide empirical shreds of evidence of our assumptions. Firstly, the reformer using the UNet has better capability to move closer to the majority of the original examples than auto-encoders. Secondly, the shift frequency domain which has zero frequency component is located at the center of an image helps the detector networks to differentiate the adversarial examples and the normal examples more accurately than the spatial domain as well as the frequency domain without using the shift function. Finally, the experiment results prove that our new defense strategy is powerful and generalized well to the various adversarial attacks.

\section{Conclusion}

We proposed the new defense methodology for defending the medical image semantic segmentation models against adversarial attacks. Our defense framework improves the prediction accuracy and decreases the power of the adversarial attacks. Furthermore, our defense framework is attack-independent. Besides, our method is independent of the target deep learning neural network models. These characteristics of our defense method lead to better generalization across the adversarial attacks.


\end{document}